# The Impact of Anionic Vacancies on the Mechanical Properties of NbC and NbN: An *ab initio* Study

P. W. Muchiri[a,c], K. K. Korir[b,c], N. W. Makau[c], G. O. Amolo[a,c]

[a]*Materials Modeling Group, School of Physics and Earth Sciences, The Technical University of Kenya, P.O. Box, 52428-00200, Nairobi, Kenya.*
[b]*Physics Department, Moi University, P.O. Box, 3900-30100, Eldoret, Kenya.*
[c]*Computational Material Science Group, Physics Department, University of Eldoret, P.O. Box, 1125-30100, Eldoret, Kenya.*

**Abstract**

The development of super-hard materials has recently focused on systems containing heavy transition metal and light main group elements. Niobium carbides and nitrides have previously been identified as potential candidates, however, the volatility of carbon and nitrogen during synthesis make them prone to formation of anionic vacancies, which have the ability of changing the electronic structure, dynamical stability and adversely affecting the mechanical properties. Here, we present *ab initio* Density Functional Theory calculations that probe the occurrence of anionic vacancies as a function of concentration, thereafter, pertinent mechanical properties are investigated. Our results showed that the presence of anionic vacancies in NbC and NbN tend to deteriorate the mechanical properties and ultimately the mechanical hardness due to vacancy softening that can be attributed to defect induced covalent to metallic bond transition. Further, it was observed that anionic vacancies in NbC tend to modify its toughness, in particular, NbC in ZB becomes brittle while NbC in WZ becomes ductile in presences of C vacancies of up to 6%. On the other hand, the toughness of NbN was found to be insensitive to defect concentration of even up to 8%. Consequently, stringent control of anionic defects during synthesis of NbC and NbN is critical for realization of the desired mechanical response that can make these materials ideal for super-hard and related applications.

**Introduction**

Increased demand for hard materials whose performance in the ultra-hard materials industry is comparable or better than that of diamond but with low cost has inspired research and development of compounds that incorporate light elements such as carbon and nitrogen with 3*d*, 4*d* or 5*d* transition metals [1-2]. Indeed, the search for materials with superior hardness and related properties has been motivated in part by drawbacks associated with traditional materials such as diamond and *c*-BN. In the wake of new applications, for example, at elevated temperature, diamond becomes unstable and reacts with iron thus not quite suitable for machining steel alloys [3] while c-BN has low hardness index thus has limited applications where extreme hardness is a prerequisite [4]. Therefore, there is a need for further exploration of design and synthesis of suitable transition metal carbides and nitrides (TMCNs) to complement conventional hard materials.

Indeed, studies have shown that TMCNs are a promising alternative towards achieving low cost, robust, and versatile hard materials [5]. The novel properties associated with these materials have been attributed to a combination of factors such as high valence electrons density of transition metals, which ensures high bulk modulus, as well as a combination of transition metals with small non-metal elements such as carbon or nitrogen, yielding short and highly directional covalent bonds ideal for extreme hardness [6-7]. These classes of materials have already found application in various areas. For example, titanium nitrides are widely used as wear resistant coatings in aerospace applications [8], while rhenium sub-nitrides have shown great promise as hard conductors suited for extreme environments [9]. On the other hand, carbides such as WC and CrC are widely used in cutting and shaping other ceramics due to their excellent properties.

Synthesis of TMCNs generally occurs at elevated temperature and pressure [10], and under such conditions, formation of defects particularly the anion (carbon/nitrogen) vacancies is favoured compared to other defects such as single cationic defects or even transition metal clusters [11]. The effects of the occurrence of such anionic defects on the mechanical properties are often overlooked though their presence in substantial concentration may influence the mechanical response of the TMCNs. For example, vacancies are known to alter chemical bonding with an adverse effect on mechanical response [12]. On the other hand, vacancies can also constrain movement of dislocation thus improving mechanical response of the material [13]. Seung-Hoon *et al* [14] and Xiao *et al* [15] ascribe more fundamental reasons of the behavior of the changes in the mechanical properties, following the introduction of vacancies, to the occupancy of the metallic *dd* bonding states, further noting that there are differences in the mechanical properties versus vacancy concentration trends in transition metal nitrides and carbides. Consequently, understanding the effect of vacancies on the mechanical response of TMCNs is paramount for the design of superior devices. For this reason, identifying and manipulating the vacancies concentration may offer a facile route towards achievement of a deeper understanding of properties of these materials, and consequently their potential applications.

Despite the substantial amount of work already focused on TMCNs, there are still open questions that are yet to be addressed. For example, the effects of anionic vacancies concentration on the mechanical properties and relevant mechanisms towards achieving better control of such defects are still unclear, yet this is necessary for tailoring these materials for mainstream applications [9-13]. NbC and NbN belong to a class of TMCNs that have attracted the attention of researchers due to their potential application in ultra-hardness and related industries. In this work, we explore the effect of vacancies on mechanical properties of NbC and NbN in rocksalt (RS), zincblende (ZB), and wurtzite (WZ) phases using density functional theory (DFT). A consideration of these three phases is expected to provide additional information of the structural dependence of their mechanical properties. In particular, the study focused on anionic vacancies formation dynamics and their effects on elastic constants and mechanical properties such as bulk modulus, shear modulus, and Young's modulus. Vickers hardness is explored, and finally materials' toughness is evaluated. Whereas our calculations are performed at ground state, while applications of NbC and NbN occurs at elevated temperatures such as in drilling, cutting and machining, DFT studies have indeed been shown to give reliable results that can still be relevant for high temperature scenarios, for example, defect diffusion due to high temperature and concentration. [16].

**Computational Details**

All calculations reported in this work were performed using *ab initio* DFT approach by solving the Kohn-Sham equation [17] as implemented in Vienna *Ab Initio* Simulation Package (VASP) [18, 19].

Generalized gradient approximation (GGA) has been used to describe the exchange correlation energy as proposed by Perdew-Burke-Ernzerhof (PBE) [20] The core-electrons were replaced with projector augmented wave (PAW) potentials and the electronic wave functions were expanded in a plane-wave basis set with cut-off energy of 500 eV. Integration over Brillouin zone was performed using Monkhorst and Pack scheme [21], where a dense 11x11x11 grid was used for RS, and ZB phases, while for WZ, 10x10x5 grid was used for both NbC and NbN, and in the case of supercells, a minimum of 30 k-points were used to ensure the convergence of total energies to less than 1 meV/atom.

Full structural optimization including atomic position, the c/a ratio, and the unit cell volume were carried out for all models with electronic convergence criteria set to $1\times10^{-8}$ eV and the forces optimized to less than $1\times10^{-6}$ eV. Elastic constants were calculated using strain-stress method, where six finite distortions are performed on a well relaxed unit cell from which the elastic tensor is determined. The elastic constants are then derived using the stress-strain relationship. Other parameters such as bulk modulus ($B$), shear modulus ($G$), Young's modulus ($E$), and Poisson's ratio were derived from the Voigt-Reuss-Hill averaging scheme [22], while Vickers hardness was estimated as described by Gao et. al. [23]. In this work, we have considered anionic vacancies concentration ranging from 1-8%, which is anticipated to have negligible distortion on the crystal upon performing structural relaxation at fixed volume and similar values have been reported in experimental work [24]. Existence of limited studies that explores the implications of lower defects (C/N vacancies) concentration in NbC and NbN is the thrust of this work, since such knowledge has the potential of leading to the development of better optimization procedures and attainment of better devices based on these materials. Dynamical stability of the defective structures was determined using phonon dispersion relation and the phonon properties computed via finite displacement method with a 2x2x2 supercell to compute the force constants and the results analyzed using the Phonopy code [37].

**Results and Discussion**

**(i) Formation of Anionic Vacancies**

Reliable vacancy formation energies are obtained by adopting larger supercells that ensure errors stemming from vacancy self-interaction are minimized. Here, we consider supercells of *3x1x1*; 24 atoms, *2x1x1*; 32 atoms, *3x2x1*; 48 atoms, *2x2x2*; 64 atoms, and *2x3x2*; 96 atoms. The preferred defect site was determined by calculating the defect formation energies for the various sites and the site with the lowest formation energy considered to be preferred and was used in the subsequent calculations, as shown in Figure

1. In order to avoid defect self-interactions, supercell sizes were optimized to ensure that sufficient separation is maintained between the defect sites yet ensuring the supercell size are computationally viable.

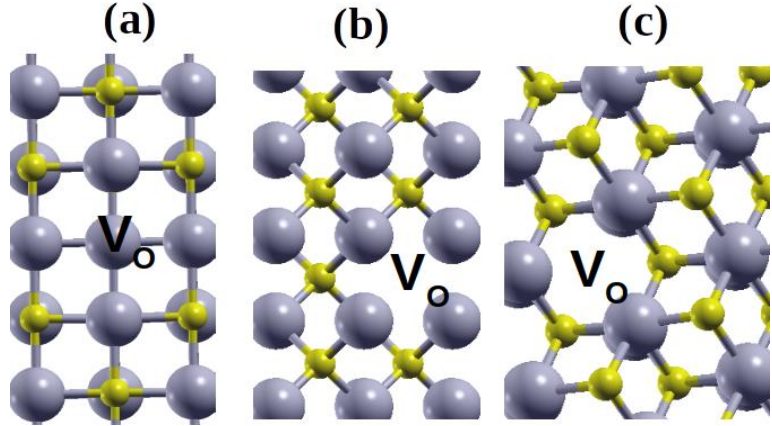

**Figure 1:** *Ball and stick representation of crystal structures of NbN and NbC considered in this work: (a) rocksalt, (b) zinc-blende and (c) wurtzite. Grey (large) spheres represent Nb atoms while the yellow (small) spheres represent either C or N atoms. Anionic vacancy site is located at $V_O$*

Using these parameters, optimization of defective structure was conducted using the optimized bulk lattice parameters as a starting point. The vacancy formation energy is calculated as shown in equation 1,

$$\Delta E_V = E_V - E_S + \mu_{C/N} \qquad (1)$$

where $E_V$ and $E_S$ are the total energies of defect-containing and stoichiometric supercells, respectively, and $\mu_{C/N}$ is the chemical potential of carbon or nitrogen [10, 11]. Generally, positive formation energies denote that the occurrence of defects can only be achieved under non-equilibrium conditions while negative formation energies imply defects can occur under equilibrium conditions.

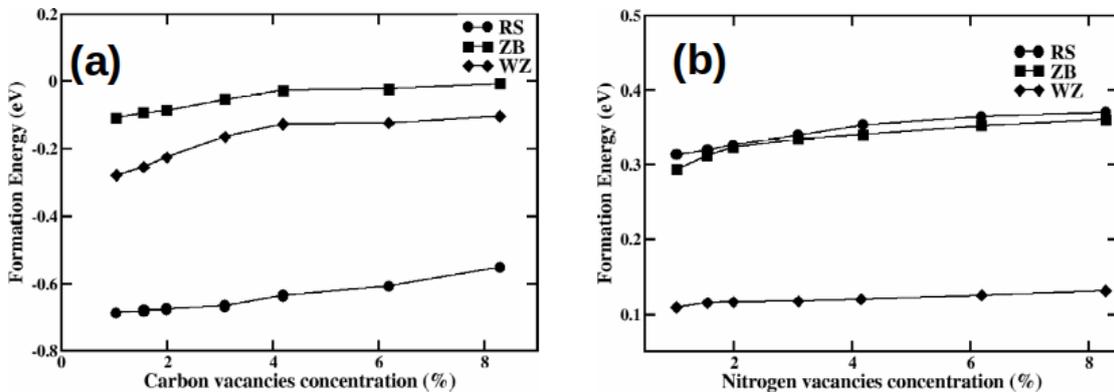

*Figure 2: Calculated formation energies for vacancies of (a) carbon in NbC (b) Nitrogen in NbN studied as a function of concentration, for RS, ZB, and WZ phases*

As shown in Figure 2(a), the calculated formation energies for carbides in RS, ZB, and WZ structures were found to be negative; indicating that formation of anionic vacancies is indeed favourable and may occur under equilibrium conditions. In carbides, the RS phase was found to favour carbon vacancies formation, compared to other phases, although, with increase in defect concentration, formation energies tend to increase (towards positive), as shown in Figure 2(a). In both ZB and WZ phases, negligible increase in formation energy is observed beyond 4 % defect concentration, while the RS phase shows an increasing trend in the formation energy even at 8% defect concentration.

On the other hand, the anionic defects in NbN tend to favour WZ phase compared to other phases considered in this work. a general observation is that with a rise in defect concentration, there is a corresponding increase of the formation energy (more positive) though mildly up to 8%, thus defect occurrence at higher concentration is not anticipated. Whereas RS and ZB phases report higher formation energies compared to WZ phase, anionic vacancy defects occurrence is generally not favoured in these phases, as shown in Figure 2(b). The rate of increase of the N vacancy formation energy in ZB appears quite marginal after 4% defect concentration and similar trend is noted is the case for RS structure beyond 6% defect concentration. This study has established that anionic defects in NbC reported negative formation energies, thus are anticipated to be more stable compared to those in NbN, where positive energies are observed for all phases considered, in agreement with previous studies [14, 15, 16, 25].

**(ii) Elastic Properties**

Generalized Hooke's law provides parameters that can be used to investigate the mechanical response of materials via the stress-strain relationship. Elastic constants are essential parameters that give insights into the relationship between crystal structure, bonding, and stability. In this work, elastic constants are determined by introducing small distortions on the cell and then relaxed via energy minimization [26]. Stress tensor with six finite distortions is utilized and the stiffness matrix analyzed using ELATE [27].

The elastics constants investigated in this work were found to be sensitive to defect concentration, and in all cases a decrease in magnitude of elastic constant is observed with increase in defect concentration. In the case of pristine structure, our calculated lattice and elastic constants were found to be consistent with those of previous studies [28, 29]. The reduction in magnitude of elastic constants can be attributed to reduction in bonding coordination number due to missing anionic atoms. In addition, vacancies in TMCNs are pinning centers which enhance the mechanical strength of these materials as they tend to inhibit dislocation motion [14]. $C_{12}$ and $C_{13}$ in RS and WZ showed a marginal decrease with increase in defect concentration (see Figures 3-5 (a-b)), which can be attributed to the non-axial nature of applied stress. These two elastic constants have lower values as compared to $C_{11}$ and therefore require small applied stress for a given strain. Thus, the non-directional stress leads to a small change in the values of these elastic constants. It is further established that, $C_{11}$ and $C_{33}$ for NbN in WZ, and $C_{12}$ for NbC in ZB showed significant reduction, for example of up to 54 % in case of $C_{11}$ for NbC in ZB phase compared to pristine value, as shown in Figure 4 (a) for defect concentrations of up to 8 %.

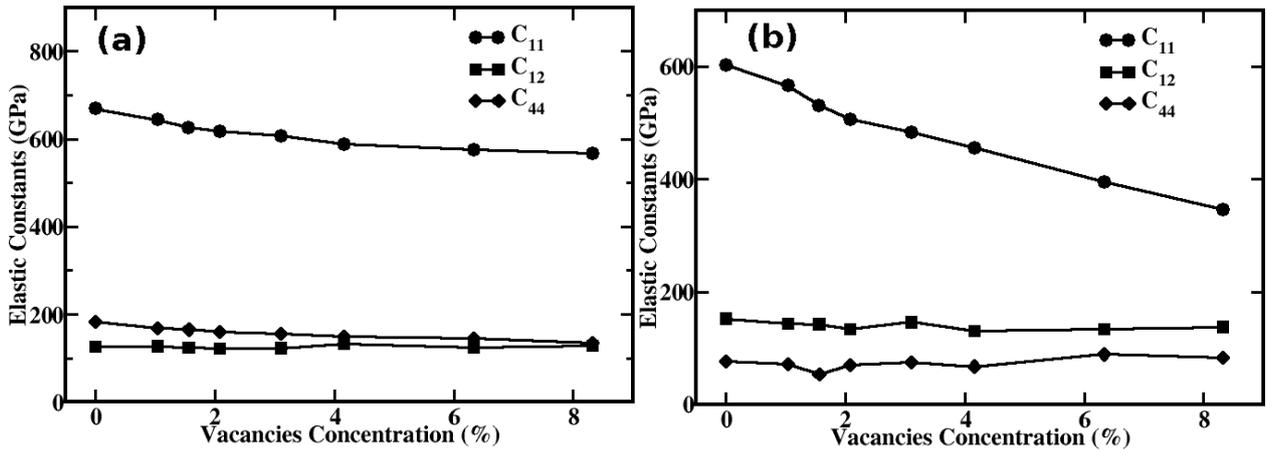

*Figure 3: Elastic constants $C_{11}$, $C_{12}$, and $C_{44}$ of RS phase (a) for NbC and (b) for NbN as a function of carbon and nitrogen vacancies concentration, respectively.*

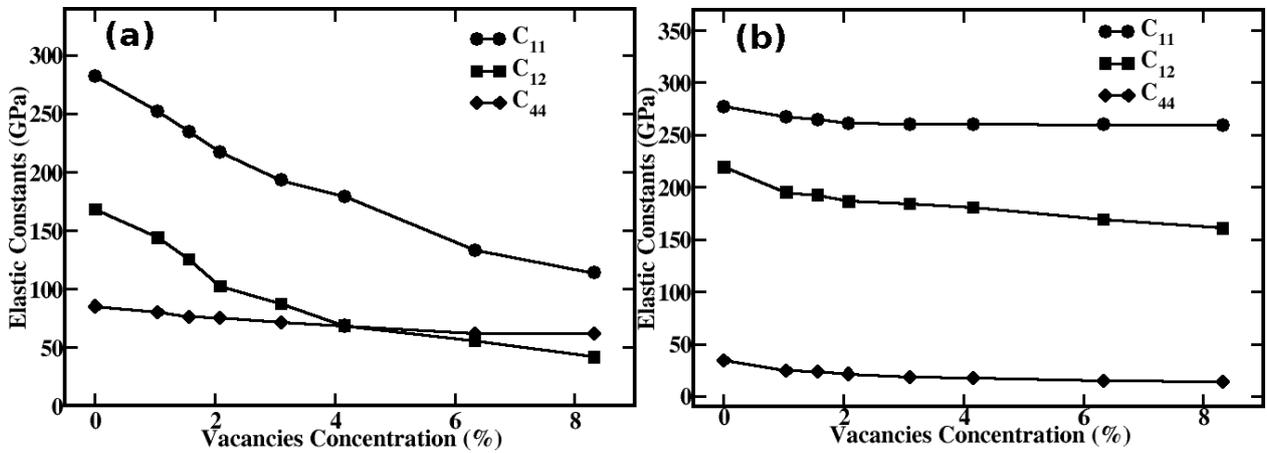

**Figure 4:** *Elastic constants $C_{11}$, $C_{12}$, and $C_{44}$ of ZB phase (a) for NbC and (b) for NbN as a function of carbon and nitrogen vacancies concentration, respectively.*

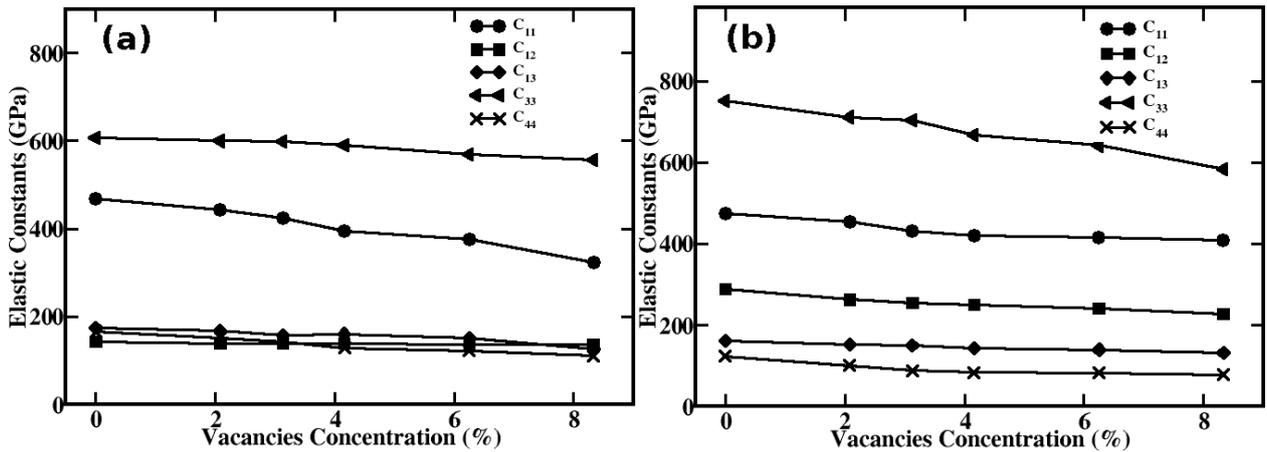

*Figure 5: Elastic constants $C_{11}$, $C_{12}$, $C_{13}$, $C_{33}$ and $C_{44}$ of WZ phase (a) for NbC and (b) for NbN as a function of carbon and nitrogen vacancies concentration, respectively.*

In RS phase, an increase in vacancies concentration leads to decrease in $C_{11}$ in both NbC and NbN, while $C_{12}$ and $C_{44}$ tend to be insensitive to changes in vacancies concentration, as shown in Figures 3(a-b). In the ZB phase, increase in defect concentration up to 8% results in corresponding decrease in $C_{11}$ and $C_{12}$ of 54 % and 57 %, respectively, while $C_{44}$ is only mildly affected with a decrease of 12 % being observed for the case of NbC, as shown in Figure 4(a). In the case of NbN, a decrease in $C_{11}$, $C_{12}$, and $C_{44}$ is observed with an increase in defect concentration of upto 4%. Beyond this concentration only a marginal effect on $C_{11}$ and $C_{11}$ elastic constants are noted, as shown in Figure 4 (b).

In the case of WZ phase, increase in the vacancies concentration tend to strongly affect $C_{11}$ with a decrease being observed in NbC, as shown in Figure 5(a), while other elastic constants remain largely unaffected. Similarly, in NbN, $C_{33}$ decreases with increase in vacancies concentration, while other elastic constants are only mildly affected, as shown in Figure 5(b).

Furthermore, it is noted that all the independent elastic constants obey the Born criterion [30], where for cubic phase (RS, ZB) $C_{44} > 0$, $C_{11} > C_{12}$ and $C_{11} + C_{12} > 0$, while in the hexagonal phase (WZ) $C_{33} > 0$, $C_{44} > 0$, $C_{12} > 0$, $C_{11} > C_{12}$, $(C_{11} + 2 C_{12}) C_{33} > 2 C^2_{13}$. This is a confirmation that these phases are indeed mechanically stable with up to 8% anionic vacancies concentration. It is generally acknowledged that anionic vacancies perturb the symmetry of the crystal hence reduced stability. Thus, stability analysis can provide insights on the mechanical stability of the crystal in presence of defects and based on our results, the phases explored were found to be mechanically stable thus confirming the possibility of the occurrence of vacancies in these structures.

**(ii) Dynamical Properties**

This study also explored dynamical stability of the various phases considered using phonon calculations, where phonon dispersion curves were obtained at arbitrary wave-vector of the Brillouin zone. If a crystal lattice with n-atoms per unit cell is considered, then *3n* branches [40]. For example, in RS phase, there are two atoms per unit cell, so there are six normal modes (3 acoustic and 3 optical). In terms of lattice dynamics, a crystal is stable if all normal vibration modes have real and finite frequencies. As shown in Figure 6, both NbC and NbN of the same phase exhibits approximately the same dispersion curve and because of lack of imaginary frequencies in NbC in RS, ZB, and WZ structures (see Figure 6 (a, c, e)), and NbN in WZ (see Figure 6 (f)) are considered stable. This is consistent with previous studies [36, 39], while NbN in RS and ZB structures are considered unstable due to presence of imaginary frequencies, as shown in Figure 6 (b, d) and in-agreement with other theoretical studies [38].

NbC (RS, ZB, and WZ structures) and NbN (WZ) are insensitive to presence of C and N vacancies, respectively, and remain dynamical stable, as shown in Figure 7 (a, c, e, f) whereas NbN in RS and ZB becomes dynamically stable with the presence of N vacancies (see Figure 7 (b, d). Thus, anionic defect at lower concentration (< 8%) tend to improve dynamical stability of NbC and NbN.

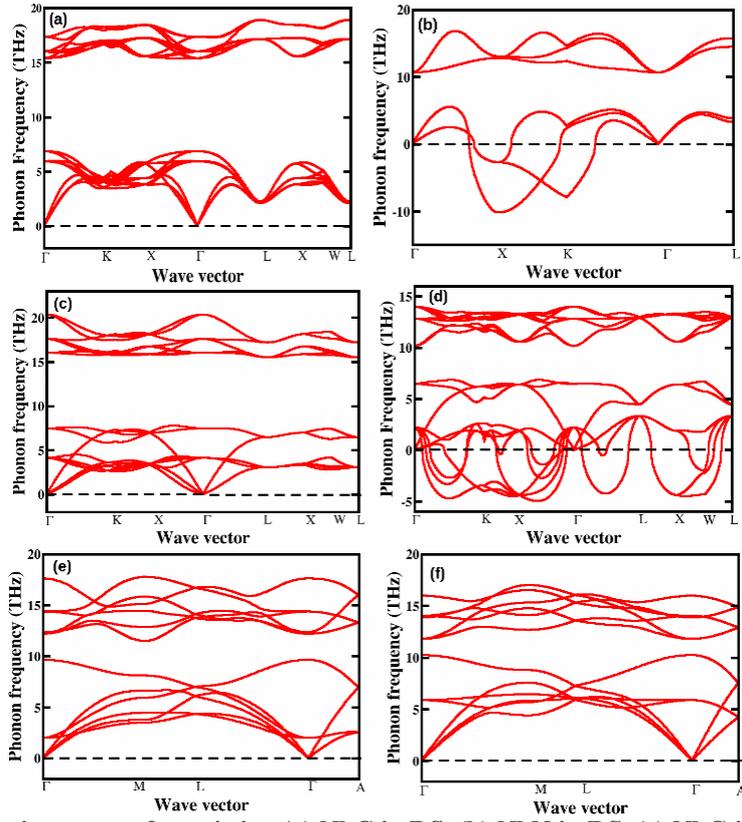

**Figure 6:** Phonon dispersion curves for pristine (a) NbC in RS, (b) NbN in RS, (c) NbC in ZB, (d) NbN in ZB, (e) NbC in WZ, and (f) NbN in WZ. Black dotted line at 0 Hz separates stable (>0) and unstable (<0) regime.

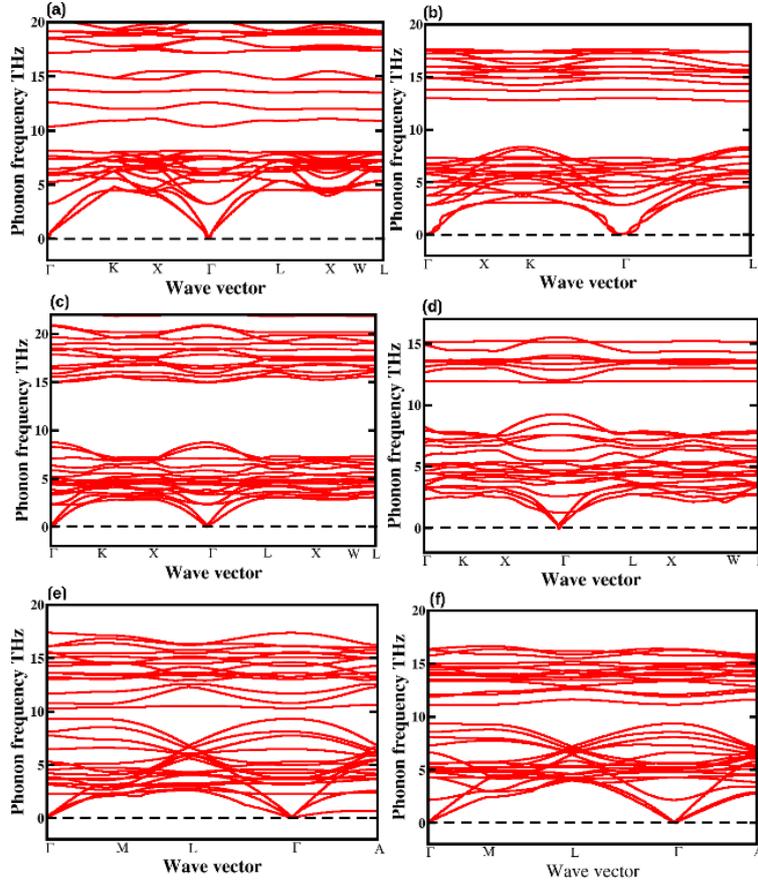

**Figure 7:** Phonon dispersion curves for defective structures with vacancies concentration of 4% for (a) NbC in RS, (b) NbN in RS, (c) NbC in ZB, (d) NbN in ZB, (e) NbC in WZ, and (f) NbN in WZ. Black dotted line at 0 Hz separates stable (>0) and unstable (<0) regime.

**(iv) Mechanical and Electronic properties**

Mechanical properties such as bulk modulus (*B*), shear modulus (*G*), and Young's modulus (*E*) are essential in predicting the applications where materials such as NbC and NbN can be used. These are determined using the Voight-Reuss-Hill approximation [31] as expressed in equations 2-4,

$$B = \frac{C_{11} + 2C_{12}}{3} \quad (2)$$

$$G = \tfrac{1}{2}(G_V + G_R) \quad (3)$$

Where $G_V = \frac{C_{11} - C_{12} + 3C_{44}}{4}$ and $G_R = \frac{2C_{44}(C_{11} - C_{12})}{4C_{44}(C_{11} - C_{12})}$

$$E = \frac{9G}{3+k} \quad (4)$$

With *k=G/B* being Pugh's ratio while $G_V$ and $G_R$ are Reuss and Voigt shear modulus, respectively.

Using the elastic constants, it is possible to predict Vickers hardness ($H_V$) of materials and despite the challenges associated with assessment of hardness, fitting functions based on the calculated *B* and *G* values have proved reliable in prediction of intrinsic hardness of a wide range of materials [32] and this can be expressed as in equation 5,

$$H_V = 2(k^2 G)^{0.585} - 3 \quad (5)$$

The computed values of the relevant mechanical properties are shown in Figure 8. This study established that increase in defect concentration tends to either increase the bulk modulus or remain unchanged in some cases for defect concentration of up to 1.5 %, thereafter, an increase in defect concentration leads to a decrease in the bulk modulus in all the structures considered in this work, as shown in Figure 8(a). Similar trend is observed in the case of Young's modulus whereby a notable decrease with increase in defect concentration is established. Beyond 4% defect concentration, the Young's modulus is however not affected by any increase in defect concentration. This means that the stiffness of the materials is barely affected after a defect concentration of 4%. In particular, the ZB structure for both NbC and NbN becomes insensitive to changes in defect concentration, thus negligible changes in the Young's modulus is noted. For the WZ phase however, a steady and almost linear decrease in Young's modulus is observed even at 8% defect concentration, as shown in Figure 8(b).

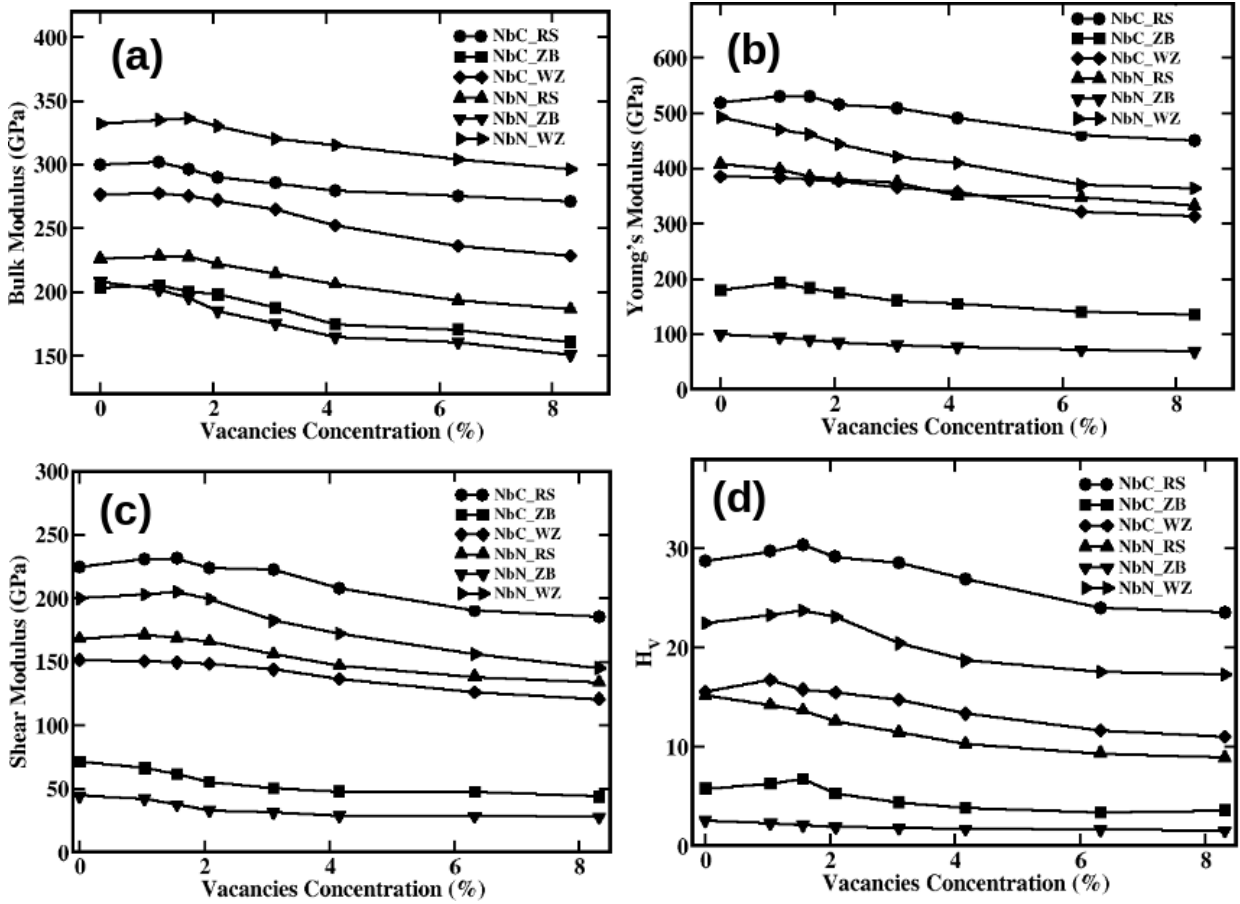

**Figure 8:** *Mechanical properties (a) bulk modulus, (b) Young's modulus, (c) Shear modulus, and (d) Vickers ($H_V$) hardness plotted as a function of vacancies concentration for Nbc and NbN in RS, ZB, and WZ phases, respectively.*

Both shear modulus and Vickers hardness showed a decrease in magnitude with increase in defect concentration beyond 1.5%, consistent with other mechanical properties considered in this work, as shown in Figure 8(c-d). The rate of decrease for both Vickers hardness and the shear modulus were not the same for NbC and NbN, a property that may be attributed to the local bonding environments in the two materials. A significant decrease in shear modulus was observed in the WZ phase, while the least was seen in the ZB phase. In the case of the NbN structure, the largest decrease in the Vickers hardness was seen in the RS phase while the ZB phase exhibited the least response. Indeed, it can be noted that all the mechanical properties considered in this work (*B, G, E,* and $H_V$) tend to decrease with increase in defect concentration. It is worth noting that experimental studies [41] on NbC and NbN in RS structure shows that in the case of shear modulus, increase in vacancy concentration ( upto 12.5 %) leads to marginal decrease in shear modulus, thereafter, it decreases monotonically with increase in vacancy concentration. Since our study considers lower vacancy concentration, we can only speculate that at higher concentration (> 12.5 %), more dramatic decrease in shear modulus and other mechanical properties considered in this work is anticipated on the strength of study being able to produce results that are consistent with other previous studies [35, 36, 41] at lower vacancy concentration. As such, our study is indeed anticipated to offer a predictive picture on the effect of anionic vacancies on the mechanical properties of NbC and NbN. In the presence of defect concentration of up to 8%, our systems of interest would still be classified as hard material (with $H_V > 0$). However, hardness is significantly deteriorated in all structures considered and they may not perform well in hardness related application. For example, NbC in RS structures loses hardness by up to 36 % at 8% defect concentration when compared to pristine structure. Therefore, we can deduce that stringent control of defect concentration in TMCNs is essential for their utilization in hardness related industries.

Toughness is another critical parameter that measures the degree of plastic deformation (ductility) of a solid under mechanical loading and can be used to predict the robustness of the materials. In this work, ductility is analyzed using the Pettifor criterion [33], whereby when the Cauchy pressure ($C_{12}$ - $C_{44}$) is positive or negative, the system is considered ductile or brittle, respectively. Also, the Pugh's modulus [34] is utilized in this analysis, with materials having *B/G > 0.57* considered brittle, while those with *B/G < 0.57* considered ductile. Thus, using both the Pettifor's and Pugh's criteria, we analyzed the effect of anionic vacancies on toughness of NbC and NbN in RS, ZB, and WZ structures. In the case of pristine structures, NbC in RS and WZ phases are found to be brittle, while NbC in ZB, and NbN in RS, ZB and WZ phase are ductile, as shown in Figure 9(a).

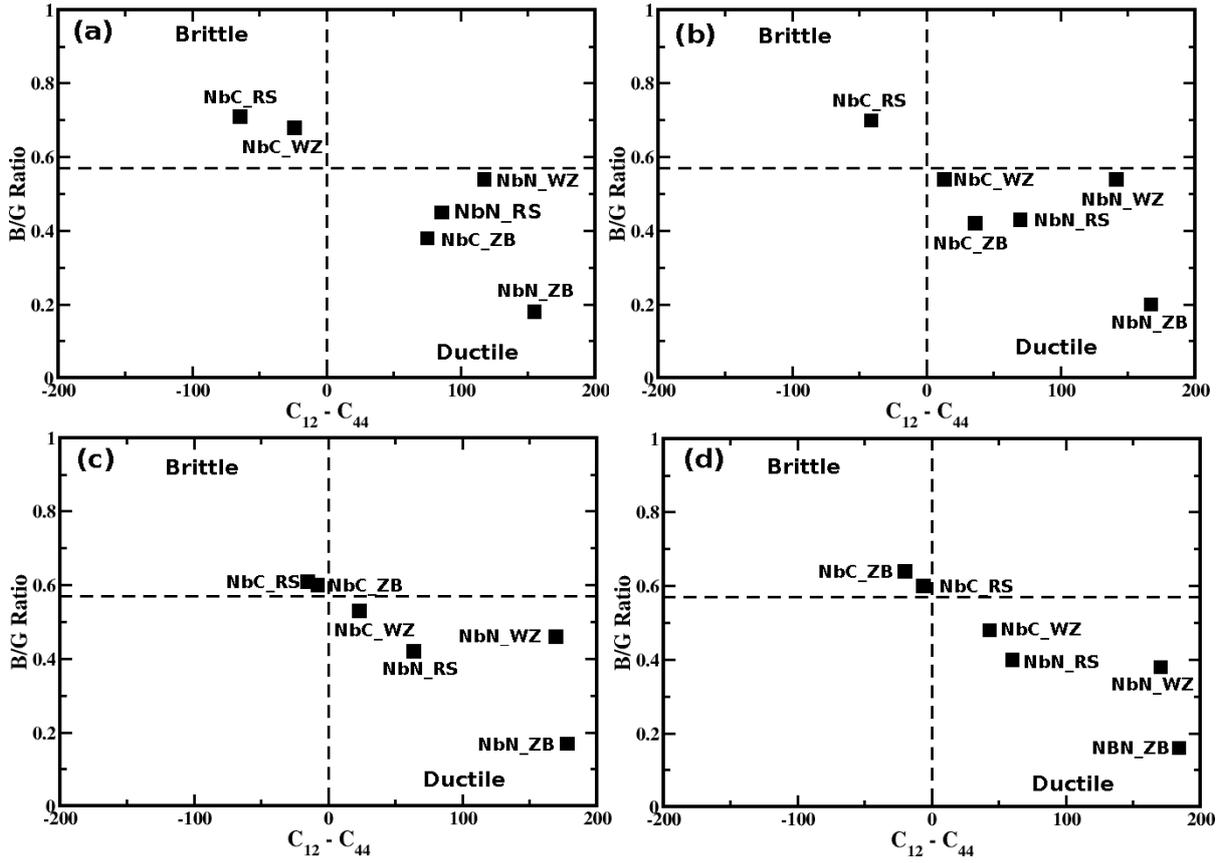

*Figure 9: Map of brittleness and ductility trends of NbC and NbN as a function of defect concentration (a) pristine, (b) 2%, (c) 4 %, and (d) 8%. B/G > 0.57 is considered brittle while B/G < 0.57 is considered ductile. Positive and negative Cauchy pressure denotes brittle and ductile materials, respectively.*

In this work, it is established that an increase in anionic vacancies concentration of upto 2% tend to reduce brittleness of NbC in RS and NbC in WZ. Moreover, NbC in WZ becomes ductile and the ductility increases with increase in vacancies concentration. NbC in RS maintains its brittleness even at 8 % vacancies concentration, but this is significantly diminished compared to pristine structure, as shown in Figure 9 (b-d). In the case of NbC in ZB structure, increase in vacancies concentration reduces its ductility and at 4 % it becomes brittle, as shown in Figure 9 (c) and this can be attributed to bonds rigidity induced by defects. Finally, NbN in RS, NbN in ZB, and NbN in WZ phases maintain their ductility and are only moderately affected by vacancies concentration of up to 8%, as shown in Figure 9 (b-d). Therefore, we can conclude that ductility of NbN is insensitive to vacancies concentration of upto 8%, while the toughness of NbC is dramatically affected by vacancies concentration. In addition, the Poisson's ratio, which is normally used to predict the material as either being ductile or brittle, was also considered. Ductility and brittleness, in this case are separated by a value of about 0.26, whereby a value of Poisson's ratio of more than (less than) 0.26 corresponds to ductility (brittleness) of the material, respectively [3].

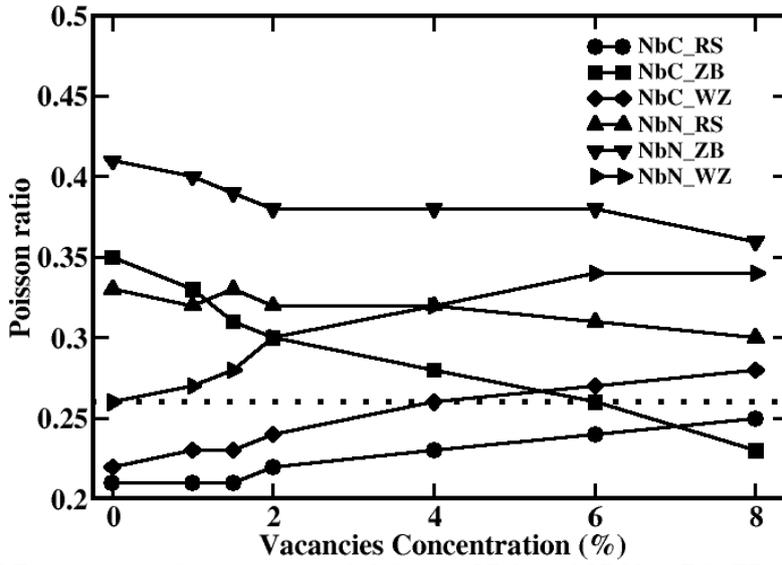

**Figure 10:** Calculated Poisson ratio for pristine and defective NbC and NbN in RS, ZB, and WZ with anionic vacancies concentration ranging from 1- 8 %. The dotted line separates brittle and ductile regions.

As shown in Figure 10, the values of the Poisson's ratio for the materials considered in this study range from 0.21 to 0.41. Based on this data, NbC in RS maintains its brittleness even with C vacancies concentration of upto 8% though it is greatly depleted deteriorated, while NbC in WZ phase becomes ductile above 6% defect concentration. On the other hand, NbC in ZB becomes brittle when C vacancies concentration increases up to 6%. The toughness of NbN in RS, ZB, and WZ considered in this work tend to be insensitive to defect concentrations of upto 8% and remain largely ductile. This picture is consistent with the Pettifor's criterion presented earlier [3] and is supported by our current study.

Changes that occur in bonding due to anionic vacancies not only affect the bond-lengths but also modify the electronic structure. In this work, we analyzed projected density of states (PDOS) of pristine and defected structures shown in Figure 11 to unravel the effect of vacancies on electronic structure. This was to help us understand the implication of C/N vacancies on bonding and ultimately on the mechanical properties, which was the thrust of the current work. It is observed that in NbC and NbN in RS phase (see Figure 11(a-b), two dominant peaks near -5eV and 4eV with reference to the shifted Fermi level at 0 can be associated with *pd* bonding between the Nb and C/N atoms, which is covalent in nature. These peaks decrease with the introduction of C/N vacancies in the structure, as shown in Figure 11 (a-b), signaling a decrease in the contribution from *pd* bonds. Similar trend is also observed for ZB and WZ phases, as shown in Figure 11(c-d), and Figure 11(e-f), respectively. Notably, in the case of ZB structure, the dominant peaks in the carbides occur at 2 eV while in nitrides it occurs at 0 eV but is significantly reduced compared to that of carbides. Thus, reduced *sp* bonding is anticipated in nitrides compared to carbides, as shown in Figure 11 (c-d). In WZ structure, introduction of C vacancies led to a dramatic decrease in the peak located at 0 eV while the peak located at 4 eV was found to increase, with similar trend being observed in nitrides, as shown in Figure 11 (e-f).

Furthermore, it is noted that there is a slight increase in states at the Fermi level with increase in C/N vacancies for all the phases considered in this work: this observation can be attributed to increase in *dd* bonding between *Nb* atoms with decrease in C/N content, thus inducing a transition from covalent to metallic bonding character of NbC and NbN. This results in reduced bonding strength with an overall reduction in magnitude of elastic constants and mechanical properties as established in this current study (see Figures 3-5 and Figure 8, respectively).

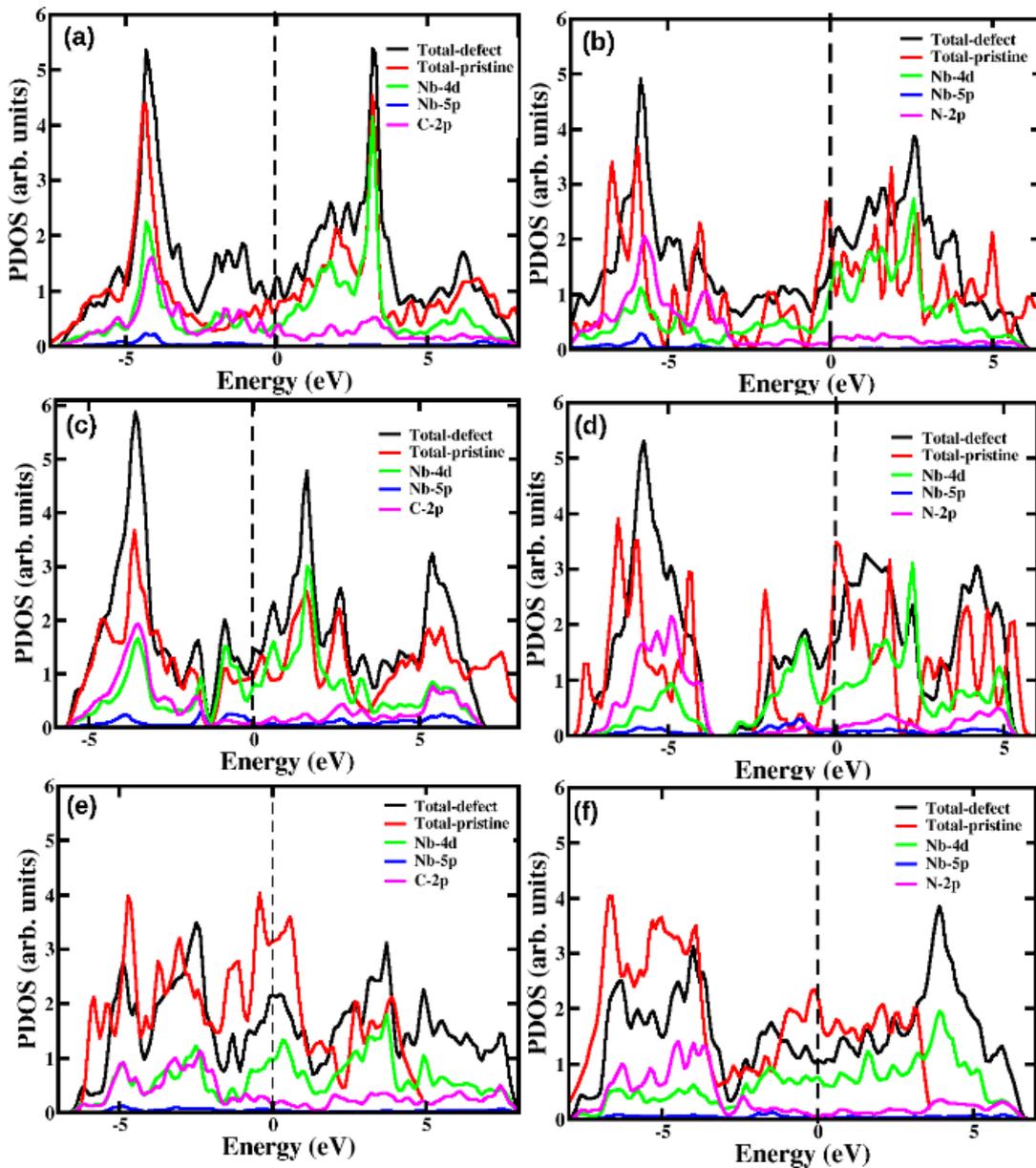

**Figure 11:** PDOS for (a) NbC in RS, (b) NbN in RS, (c) NbC in ZB, (d) NbN in ZB, (e) NbC in WZ, and (f) NbN in WZ. Dotted line at 0 eV is the Fermi level and defect concentration of 8 % is considered.

**Conclusions**

In summary, this study has investigated in detail through ab initio DFT simulations the effect of anionic vacancies on the mechanical, elastic and some electronic properties of NbC and NbN in RS, ZB and WZ structures as well as their dynamical stability. The formation of low concentrations of anionic defects was found to be favoured in NbC, while their occurrence in NbN can only be achieved under non-equilibrium conditions. The phonon dispersions curves of pristine niobium carbide and nitride showed that NbC (in RS, ZB, and WZ structures) with NbN in WZ structure are dynamically stable while NbN (in RS and ZB structures) are dynamically unstable. Additionally, NbC (in RS, ZB, and WZ structures) and NbN (in WZ) are insensitive to presence of C and N vacancies, respectively, and remain dynamically stable whereas NbN in RS and ZB becomes dynamically stable with the presence of N vacancies.

Our results showed that the presence of C and N vacancies in NbC and NbN, respectively leads to an overall depletion in bulk modulus, shear modulus, Young's modulus, elastic constants, and Vickers hardness. For instance, the Vickers hardness of NbC in RS is reported to have been reduced by 36% in the presence of anionic vacancies at 8% defect concentration. The deterioration of the mechanical properties can be attributed to vacancy softening due to decrease in the number of covalent bonds and increased occupation of anti-bonding states. The states at the Fermi level tend to increase slightly with increase in C/N vacancies and these can be associated with rise in *dd* bonding states between Nb atoms with decrease in C/N content. Further, it was observed that anionic vacancies in NbC tend to modify its toughness; in particular, NbC in ZB becomes brittle while NbC in WZ becomes ductile in the presences of C vacancies of up to 6%. The toughness of NbN was found to be generally insensitive to defect concentrations of upto 8%. Therefore, stringent control of anionic defects in NbC and NbN is critical for their optimal mechanical response.

**Acknowledgments**
The authors acknowledge funding from ACEII-PTRE of Moi University, and CHPC-Cape Town for the availability of High-Performance Computing resources and support.